\documentclass[aps, prl, superscriptaddress, twocolumn,10pt]{revtex4-1}
\usepackage[left=1in, right=1in, top=1in, bottom=1in]{geometry}

\usepackage{bm}
\usepackage{tikz,pgfplots}
\usepackage{standalone}
\usepackage[retainorgcmds]{IEEEtrantools}
\usepackage{graphicx}
\usepackage{mathrsfs}
\usepackage{amsmath}
\usepackage{amssymb}
\usepackage{color,libertine}
\usepackage{amsfonts}
\usepackage{times,txfonts}
\usepackage{nicefrac}
\usepackage[colorlinks=true,linkcolor=blue,urlcolor=blue,citecolor=blue,pdfusetitle]{hyperref}
\usepackage{hyperref}
\usepackage[normalem]{ulem}
\usepackage{calrsfs}
\DeclareMathAlphabet{\mathcal}{OMS}{cmsy}{m}{n}
\usepackage{amssymb}
\usepackage{natbib}
\newcommand{\ket}[1]{\left| {#1} \right\rangle}
\newcommand{\bra}[1]{\left\langle {#1}\right|}

\definecolor{ao}{rgb}{0.0, 0.5, 0.0}
\definecolor{mypink}{rgb}{0.858, 0.188, 0.478}
\definecolor{mygreen}{rgb}{0.0, 0.5, 0.0}

\begin{document}

\title{The role of quantum coherence in energy fluctuations}

\author{S. Gherardini}\thanks{These authors contributed equally to this work}
\affiliation{\mbox{Department of Physics and Astronomy \& LENS, University of Florence,} via G. Sansone 1, I-50019 Sesto Fiorentino, Italy.}
\affiliation{CNR-IOM DEMOCRITOS Simulation Center and SISSA, Via Bonomea 265, I-34136 Trieste, Italy}
\author{A. Belenchia}\thanks{These authors contributed equally to this work}
\affiliation{Centre for Theoretical Atomic, Molecular and Optical Physics, School of Mathematics and Physics, Queen's University Belfast, Belfast BT7 1NN, United Kingdom}
\author{M. Paternostro}
\affiliation{Centre for Theoretical Atomic, Molecular and Optical Physics, School of Mathematics and Physics, Queen's University Belfast, Belfast BT7 1NN, United Kingdom}
\author{A. Trombettoni}
\affiliation{Department of Physics, University of Trieste, Strada Costiera 11, I-34151 Trieste, Italy}
\affiliation{CNR-IOM DEMOCRITOS Simulation Center and SISSA, Via Bonomea 265, I-34136 Trieste, Italy}

\begin{abstract}
  We discuss the role of quantum coherence in the energy
  fluctuations of open quantum systems. To this aim,
  we introduce an operational protocol, to which we refer to
  as the end-point-measurement scheme, 
  allowing to define the statistics of energy changes as a function
  of energy measurements performed only after its evolution.
  At the price of an additional uncertainty on the value of
  the initial energies, this approach prevents the loss of initial
  quantum coherences and enables the 
  estimation of their effects on energy fluctuations. 
  We illustrate our findings using a three-level quantum system in
  interaction with thermal reservoirs.
\end{abstract}

\maketitle

When the size of a physical system is scaled down to enter the
micro-/nano-scopic domain, the fluctuations of relevant physical quantities
start playing a pivotal role in establishing the energetics
of the system itself. Such fluctuation obey fundamental relations,
going under the name of \textit{fluctuation theorems}, that recast
the laws of thermodynamics in such a new regime of operation.
The definition of familiar thermodynamic quantities should be
refined at such micro- and nano-scales to
account for fluctuation-induced physical effects.
Should the range of energies involved in a given system bring
its dynamics within the domain of quantum theory, the very nature of
such fluctuations become even more interesting as encompassing
both classical -- i.e., thermal -- and quantum contributions.
The characterization of the latter, and the understanding of their
interplay with the former, so as to set the dynamics of fundamental
energy transformations at the quantum level, are daunting
yet very stimulating open problems. 

One of the key achievements of the emerging field of
thermodynamics of quantum
processes~\cite{Vinjanampathy2016,Sagawa2014,BookQT,deffner2019qtd}
is the identification of a strategy for the assessment of the
energetics stemming from non-equilibrium quantum dynamics.
The so-called two-point measurement (TPM)
protocol~\cite{TalknerPRE2007,EspositoRMP2009,CampisiRMP2011}, where
the energy is measured both at the initial and final time, has been
introduced with the purpose of determining the work statistics
of a quantum system driven by a time-dependent protocol. 

The main idea behind this protocol stems from classical considerations:
The energy-change of a given system is determined by measuring energy before
and after the dynamics takes place. Such evaluation depends 
only on the knowledge of the Hamiltonian that drives
the process and not on the procedure or apparatus used to measure it.
However, in quantum mechanics, measurements play an
active role in that they condition the dynamical evolution of a
system~\cite{BookJacobs}. In particular, in the TPM protocol the first
energy measurement -- performed before the dynamics takes place --
destroys the quantum coherences
(and possible quantum correlations with the environment)
in the initial state of the system, forcing the system into an energy
eigenstate~\cite{AllahverdyanPRE2014,LostaglioNATCOMM2015}. Such a loss of
coherence is common to interferometric formulations of TPM,
which have been put forward to ease the experimental inference
of the statistics of non-equilibrium thermodynamic
quantities~\cite{Mazzola2013,Dorner2013,Batalhao2014}.

Recently, much effort has been devoted to understand the role of
coherence in quantum thermodynamics~\cite{SolinasPRE2015,SolinasPRA2016,AlhambraPRX2016,AbergPRX2018,LostaglioPRL2018,XuPRA2018,FrancicaPRE2019,SantosnpjQI2019,MingoQuantum2019,MicadeiPRL2020}.
In particular in
Refs.~\cite{SolinasPRE2015,SolinasPRA2016,XuPRA2018,LevyArxiv2019}
full counting statistics~\cite{NazarovEURPHYS2003,ClerkPRA2011}
has been put in place to study work fluctuations
in quantum systems initialized in an arbitrary state,
pointing out that the quantum interference stemming from
solely taking into account quantum coherence terms
could lead to negative quasi-probability work distributions~\cite{HoferPRL2016}.

In this paper, we propose an
\textit{operational end-point-measurement (EPM)} protocol to
quantify the statistical moments of energy fluctuations
in the (possible) presence of quantum coherence in the initial
state of the quantum system. Such a protocol removes
the need for the first projective measurement required in the TPM protocol,
thus preventing the collapse of the initial state of the system onto
the energy basis.
This is in contrast with recent proposals such as the one given
in Ref.~\cite{MicadeiPRL2020}, where the system has to be
initialized in a mixture of eigenstates pertaining to an
observable $O$ that does not commute
with the system Hamiltonian. In this scheme the initial density matrix
is diagonal
in the eigenbasis of $O$, and this is equivalent
in an experimental realization to measure $O$ at the
initial time, so that in each trajectory the starting point is an eigenstate
of $O$. Our proposal is different from this
(as discussed below) and other TPM schemes,
since we do not foresee any initial projective measurement
and the initial state fully evolves with the typical
interference phenomena of quantum dynamics~\cite{vonNeumann1955}.

Remarkably, our formalism is able to fully characterize the
fluctuations of energy changes by distinguishing between contributions
stemming from genuine quantum coherences and those resulting from initial
populations (i.e., the diagonal elements of the initial density matrix)
-- albeit at the cost of a quantifiable extra uncertainty.
Therefore, these results offer the possibility to single out
the thermodynamic features resulting from  coherence and
correlation-induced quantum effects, and set them apart from those
due to thermal fluctuations. As the giving away of the initial
energy measurement on the system is likely to entail a
substantive practical simplification, we expect our EPM protocol
to be viable in a variety of experimental situations,
thus promoting our protocol to be a fully fledged alternative
to the celebrated TPM scheme when quantum signatures are taken into account.

\noindent
{\bf Coherence in the energy eigenbasis.--} Let us consider a
$d$-dimensional quantum system $\mathcal{S}$ evolving
according to a one-parameter family of completely-positive and
trace-preserving (CPTP) maps
$\Phi_t:\rho_{\rm i}\rightarrow\rho_{\rm f}=\Phi_t[\rho_{\rm i}]$~\cite{Caruso_RevModPhys_2014} within the time interval
$\mathcal{I} \equiv [t_{\rm i},t_{\rm f}]$. Here $\rho_{\rm i}$ ($\rho_{\rm f}$)
is the initial (final) density operators of the system.
This general setting includes several scenarios: our derivation
can be specialized to the case of closed systems dynamics with
time-dependent Hamiltonian, where energy fluctuations just identify
as work, or to an open quantum system with a time-independent Hamiltonian,
where only heat-transfer can occur. 

In what follows, we consider the case where the system is not subject to
\textit{any} initial projective measurement
and aim at characterizing the fluctuations of
the energy only by the means of a final-time measurement.
This is
different from the TPM protocol and 
also
from Ref.~\cite{MicadeiPRL2020} where the elements of an
ensemble of identical systems should be prepared each in one of the
eigenstates of $\rho_{\rm i}$ or of an observable not commuting
with the Hamiltonian.
The only projective energy measurement of our EPM protocol is performed
at the final time instant $t_{\rm f}$, i.e., after the evolution is complete.
This approach gives rise to the dynamical trajectories
${\cal T}^k_i:\,\rho_{\rm i}\rightarrow\Pi_{\rm f}^{k}$,
with $\Pi_{\rm f}^{k} \equiv |E^{k}_{\rm f}\rangle\!\langle E^{k}_{\rm f}|$
the projector onto the $k$-th energy eigenstates $\vert E^k_{\rm f}\rangle$
of the system Hamiltonian at time $t_{\rm f}$.
The stochastic nature of the outcomes of the end-point energy measurement
with respect to the initial energies the system would have, \textit{if}
the energy had been measured, 
make the energy differences $\Delta E \equiv E_{\rm f} - E_{\rm i}$ a
random variable. 

From a dynamical viewpoint, the initial quantum coherence in the state of
${\cal S}$ in the energy basis can be taken into account by considering
the probability distribution of the final energy values dictated by
the evolved initial state $\rho_{\rm i}$, comprising its coherence.
By fixing the final energy of $\mathcal{S}$ at $t_{\rm f}$,
there is always a probability law weighting the trajectories
${\cal T}^k_i$, which can be arranged in $N$ groups
corresponding to the number of possible energy values at $t_{\rm i}$.
Such probability law has a purely classical nature and can be interpreted
as the uncertainty on the values of $E_{\rm i}$, and thus $\Delta E$.

By just performing energy measurements at the final time $t_{\rm f}$, one can
thus embed the effects of initial coherences into single realizations
of the system evolution. The uncertainty on $E_{\rm i}$ reflects
the fact that its values are obtained as if we were performing a
\textit{virtual} projective measurements, thus without effectively
considering any state collapse. This justifies the statistical
independence of the energy projective measurements at $t_\text{f}$ with
respect to the initial virtual one.

We pause here to comment about the initial state $\rho_{\rm i}$.
Suppose that it is not diagonal in the energy basis:
one can object in this case 
that it always exists an observable, let denote it by $O$,
in whose basis $\rho_{\rm i}$ is diagonal.
However, there is an expected difference between the case where \textit{a)}
a measurement of $O$ is done at time $t_{\rm i}$ and one starts each trajectory from an eigenstate of $O$ (as in Ref.~\cite{MicadeiPRL2020})
and the one where or \textit{b)} no measurement is implemented and the quantum dynamics can fully show
interference among paths. Such difference will be quantified later.

\begin{figure}[t]
\centering
\includegraphics[width=0.48\textwidth]{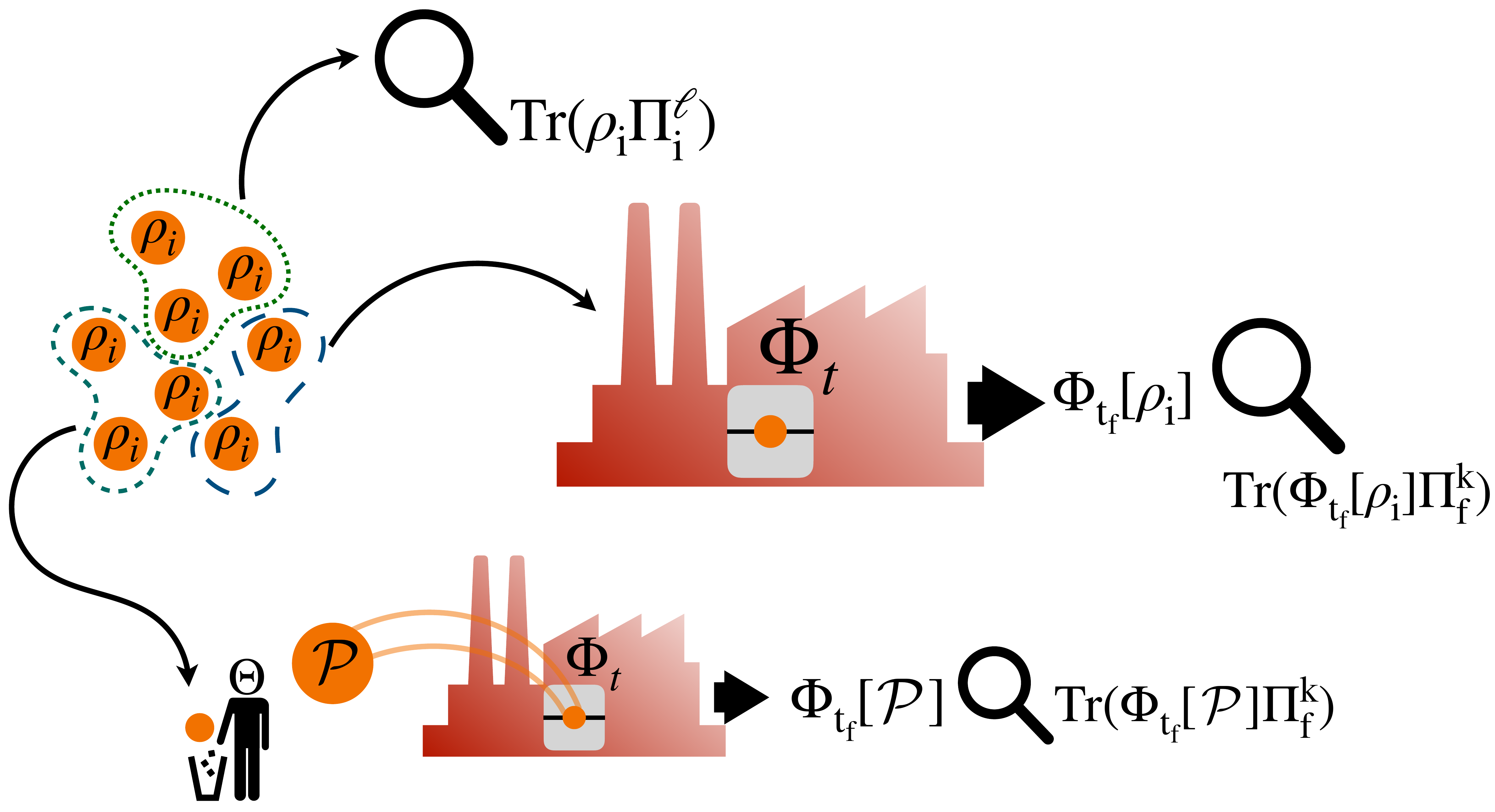}
\caption{Illustration of our operational protocol
  for the quantification of energy fluctuations and
  the extraction of information about coherence.
  An ensemble of identical systems, all prepared in the same
  initial state $\rho_{\rm{i}}$, is initially divided in three
  (in general hetero-dimensional) subgroups.
  One subgroup is used to obtain
  $p^{\ell}_{\rm i}=\rm{Tr}(\rho_i\Pi^{(\ell)}_{\rm i})$
  via an initial energy measurement. The second subgroup goes
  through a dephasing channel that returns the diagonal state $\mathcal{P}$ 
  in the energy basis. Then, $\mathcal{P}$
  is subject to the dynamical quantum map $\Phi_t$ and used to derive
  $p_{\mathcal{P}}^k=\rm{Tr}(\Phi_t[\mathcal{P}]\Pi^{(k)}_{\rm f})$
  (note that also the first subgroup, after the energy measurement,
  can be used for such purpose). Finally, the third subgroup of
  systems are those that are not initially measured but directly
  subjected to the system dynamics. These are used to obtain
  $p_{\rm f}^{k}=\rm{Tr}(\Phi_t[\rho_i]\Pi^{(k)}_{\rm{f}})$.}
\label{fig:figprotocol}
\end{figure}

Another comment is due on the initial energies $E_{\rm i}$:
if the energy is not measured at $t=t_{\rm i}$, how we can talk about them?
The point is that this information, and the related thermodynamic cost, is
encoded in the preparation of the initial state. So $\rho_{\rm i}$ is prepared
in a way that, \textit{if} we decide to measure the energy, we would find
the initial energies $E_{\rm i}$. One can think, as illustrated in
Fig.~\ref{fig:figprotocol}, that one prepares
the state $\rho_{\rm i}$ a certain (very large) number of times
and in a (finite) fraction of them one measures the energy to
verify that the $E_{\rm i}^{\ell}$ (eigenvalues of the Hamiltonian at time
$t=t_{\rm i}$) are obtained
with the probability assigned by the density matrix $\rho_{\rm i}$ -- and the
remaining times one uses the $\rho_{\rm i}$ as input for our protocol
\textit{without} measuring the energy at the time $t=t_{\rm i}$.

\noindent
{\bf Energy-change distribution and link with fluctuation relations.--}
Let us assume a time-dependent Hamiltonian process and define the
probability distribution associated to $\Delta E$ and analyze its properties.
At the single-trajectory level, the density operator after the
end-point energy measurement is one of the eigenstates $\Pi_{\rm f}^{k}$
of the time-dependent Hamiltonian $H(t_{\rm f})$. Such state is achieved
with probability
\begin{equation}\label{p_f_k}
p_{\rm f}^{k} \equiv {\rm Tr}\left(\rho_{\rm f}\Pi_{\rm f}^{k}\right) = {\rm Tr}\left(\Phi_{t_{\rm f}}[\rho_{\rm i}]\Pi_{\rm f}^{k}\right).
\end{equation}
Thus, given the energy variation
$\Delta E^{k,\ell} \equiv E^{k}_{\rm f} - E^{\ell}_{\rm i}$ in terms of the
eigenvalues of $H(t)$, the probability distribution of $\Delta E$
is obtained as
\begin{equation}\label{prob_distr}
{\rm P}_{\rm coh}(\Delta E) = \sum_{k}p_{\rm f}^{k}\sum_{\ell}p_{\rm i}^{\ell}\delta(\Delta E - \Delta E^{k,\ell}),
\end{equation}
where $p_{\rm i}^{\ell} \equiv p(E_{\rm i}^{\ell}) = {\rm Tr}
(\rho_{\rm i}\Pi_{\rm i}^{\ell})$ is the probability of obtaining
$E^{\ell}_{\rm i}$ \textit{if} an energy measurement was performed
on $\mathcal{S}$ (initial virtual measurement in the sense before specified).
In Eq.~\eqref{prob_distr}, the suffix "\rm{coh}" stands for "coherence".
The joint probability $p(E_{\rm i}^{\ell},E_{\rm f}^{k})$ associated to the
stochastic variable $\Delta E^{k,\ell}$, such that
${\rm P}_{\rm coh}(\Delta E) = \sum_{\ell,k} p(E_{\rm i}^{\ell},E_{\rm f}^{k})
\delta(\Delta E - \Delta E^{k,\ell})$,  
can be then cast into the form
\begin{equation}\label{pdf}
  p(E_{\rm i}^{\ell},E_{\rm f}^{k})= p_{\rm i}^{\ell}p_{\rm f}^{k} =
  {\rm Tr}\left(\rho_{\rm i}\Pi_{\rm i}^{\ell}\right){\rm Tr}\left(\Phi_{t_{\rm f}}
  [\rho_{\rm i}]\Pi_{\rm f}^{k}\right) \equiv p_{\rm coh}^{\ell,k}.
\end{equation}
As already noticed, the assumption behind this expression is the
statistical independence of the results of the final energy projective
measurement and initial virtual one. This comes intuitively from the
fact that the initial measurement is not performed and only
the statistics related to the initial state preparation is used.

The following properties hold for the distribution $\text{P}_{\rm coh}(\Delta E)$: \\
\noindent
    {\bf Property (i)}\label{uno} ${\rm P}_{\rm coh}(\Delta E)$
    is a probability distribution, such that $\sum_{k,\ell}p_{\rm coh}^{\ell,k}=1$. \\
\noindent
    {\bf Property (ii)} The average energy variation
    $\langle\Delta E\rangle_{{\rm P}_{\rm coh}} \equiv \int d\Delta E~{\rm P}_{\rm coh}(\Delta E)\Delta E$, correctly reproduces the expected definition
    of the average energy change induced by the CPTP map $\Phi_t$, that is
\begin{equation}\label{average_DeltaE}
  \langle\Delta E\rangle = {\rm Tr}(\mathcal{H}(t_{\rm f})\rho_{\rm f}) - {\rm Tr}(\mathcal{H}(t_{\rm i})\rho_{\rm i}),
\end{equation}
where we have used the hypothesis of statistical independence
between the final energy measurement and the virtual initial one~\footnote{
  Let us observe that, in order to obtain Eq.~\eqref{average_DeltaE}, we need to
  weight the statistics of the measurement outcomes at $t=t_{\rm f}$
  with the probabilities to initially get one of the outcomes $E_{\rm i}$.
  Otherwise the energy variation $\Delta E$ is erroneously proportional to
  ${\rm Tr}[H(t_{\rm f})\rho_{\rm f}]$.}. \\
\noindent
    {\bf Property (iii)} ${\rm P}_{\rm coh}(\Delta E)$ cannot result from a fluctuation theorem (FT) protocol.\\
\noindent    
Even by substituting a state diagonal in the (initial) energy eigenbasis
in place of the initial density operator $\rho_{\rm i}$ in
Eq.~\eqref{prob_distr}, it is not possible to directly recover
the conventional energy-change statistics resulting from the TPM protocol.
The latter is recovered only when the initial state is an
eigenstate of the energy (in this regard, see the appendix).
In this case, the discrepancy between the two joint probabilities
has to be ascribed entirely to a classical uncertainty on the initial
state of the system, which is retained in our scheme while is lost
in the TPM protocol due to the initial energy measurement.
As pointed out in the appendix, this result is in agreement
with the no-go theorem put forward in Ref.~\cite{Perarnau-LlobetPRL2017}.
As a consequence, the scheme leading to the expression of
${\rm P}_{\rm coh}(\Delta E)$ cannot be defined as a
\textit{FT protocol}~\cite{LostaglioPRL2018}. For the same reasons,
besides a few exceptions discussed in the appendix, the distribution
${\rm P}_{\rm coh}(\Delta E)$ may not be convex
under a linear mixture of protocols that only differ
by the initial density operator $\rho_{\rm i}$. This means that,
in general, given the initial density operator
$\rho_{\rm i} = \zeta\rho_{\rm i, 1} + (1-\zeta)\rho_{\rm i, 2}$
with $\zeta\in[0,1]$, ${\rm P}_{\rm coh}(\Delta E|\rho_{\rm i})$
having $\rho_{\rm i}$ as initial state cannot be expressed as a
linear composition of the distributions
${\rm P}_{\rm coh}(\Delta E|\rho_{\rm i, 1})$
and ${\rm P}_{\rm coh}(\Delta E|\rho_{\rm i, 2})$.

In order to properly single out the effect of the initial state
coherence in the energy basis, and clearly separate it from
the effects of classical uncertainty, we split the initial state
of ${\cal S}$ as $\rho_i=\mathcal{P}+\chi$, where $\mathcal{P}$
is diagonal in the energy basis while $\chi$ encodes
the coherence contributions and it is such that $\rm Tr(\chi)=0$.
Then $p_{\rm coh}^{\ell,k}$ in Eq.~\eqref{pdf} can be correspondingly split as 
\begin{equation}
\label{splittato}
    p_{\rm coh}^{\ell,k} = p_{\rm i}^{\ell}p_{\rm f}^{k} \equiv p^{\ell}_{\rm i}p_{\mathcal{P}}^k+p^{\ell}_{\rm i}p_{\chi}^{k},
\end{equation}
with
\begin{equation}
\label{splittato2}
p_{\rm f}^{k} \equiv p_{\mathcal{P}}^k+p_{\chi}^{k}={\rm Tr}(\Phi_t[\mathcal{P}]\Pi^{k}_{\rm f})+{\rm Tr}(\Phi_t[\chi]\Pi^{k}_{\rm f}).
\end{equation}

The first term, $p^{\ell}_{\rm i}p_{\mathcal{P}}^k$, in Eq.~\eqref{splittato}
encodes information on
classical uncertainty on the initial system populations, while the
second one, $p^{\ell}_{\rm i}p_{\chi}^{k}$,
takes into account the effects of initial coherence. In the following
the notation $p_{\rm coh}^{\mathcal{P}} \equiv p^{\ell}_{\rm i} p_{\mathcal{P}}^k$
will be used.
Owing to the statistical independence of outcomes $\{E^{\ell}_{\rm i}\}$
and $\{E^{k}_{\rm f}\}$, such terms can be separately analyzed.
The term~\eqref{splittato2} containing the information on the initial
coherence can be experimentally determined as illustrated in
Fig.~\ref{fig:figprotocol}, where we discuss how to obtain $p_{\rm i}^{\ell}$,
$p_{\mathcal{P}}^k$ and $p_{\rm f}^{k}$. Using Eq.~\eqref{splittato} one can
determine $p_{\chi}^{k}$.

It is worth pointing out that in the absence of initial coherences
Eq.~\eqref{pdf} is equivalent to the product of the marginals of the
probability distribution of the TMP
scheme~\footnote{Similarly, the same result holds if we compare the probability density function of the EPM protocol, for a general initial state this time,
  with the one of the MLL scheme~\cite{MicadeiPRL2020}
  (see also the SM to this work). We thank Gabriel Landi for
  pointing out this result in relation to the MLL scheme.}.
We thus have $H(p_{\rm{TPM}})\leq H(p_{\rm{coh}}|_{\chi=0}),$ which follows
from the positivity of mutual information (here $H(p)$ stands for the Shannon entropy aof a given distribution $p$). However, the same result
is not true in general if initial coherence is present.

Before going further, let us address the differences with
the protocol put forward in Ref.~\cite{MicadeiPRL2020} -- labelled as
MLL from here on -- to study the effects of coherence on heat fluctuations.
In such a scheme, a general initial state, decomposed in terms of its
eigenstates $\{\vert s\rangle\}$ as $\rho_i=\sum_s p^{s}|s\rangle\!\langle s|$,
is associated with the joint probability
$p_{\rm MLL}^{\ell,k}\equiv\sum_s p^{s} |\langle s|E^{\ell}_{\rm i}\rangle|^2\rm{Tr}(\Phi_{t_{\rm f}}[|s\rangle\!\langle s|]\Pi^{k}_{\rm{f}})$.
The latter reduces to the joint probability of the TPM protocol
for an initial state diagonal in the energy basis and to the distribution
$p_{\rm coh}^{(\ell,k)}$ in our protocol for any initial pure state. However,
for a generic initial state, such correspondences are lost and the protocol
in Ref.~\cite{MicadeiPRL2020} requires $\rho_i$ to be initialized in its
own eigenstates, corresponding to a projective measurement on the eigenstates
basis. In fact, the construction of $p_{\rm{MLL}}^{\ell,k}$ requires the knowledge
of the evolution of each individual components of $\rho_i$. In this regard,
our protocol requires less information on the system dynamics -- but at the
cost of an extra uncertainty on the statistics of $\Delta E$
(cf. the appendix). A detailed discussion of the comparison betweem EPM, MLL and TPM protocols is in the appendix.

\noindent
    {\bf Linear response approximation.--}We now further characterize
    the distribution of energy changes and address its $1^\text{st}$
    and $2^\text{nd}$ statistical moments in comparison with the
    corresponding quantities achieved using some of the
    other protocols mentioned above. As it occurs when using the
    MLL protocol, our Eq.~\eqref{average_DeltaE} recovers
    the expected difference of the averaged initial and final Hamiltonian.
    However, this is true for the TPM protocol only when the initial state
    of the scheme is taken to be the mixture resulting from the first
    energy measurement.

    As for the $2^\text{nd}$ moment $\langle\Delta E^2\rangle$,
    which accounts for the fluctuations of the random variable
    $\Delta E$ under the linear response approximation,
    from Eq.~\eqref{prob_distr} one gets
\begin{equation}\label{second_mom}
\begin{aligned}
 \langle\Delta E^2\rangle &= \,{\rm Tr}({H}^2(t_{\rm i})\rho_{\rm i}) + {\rm Tr}({H}^2(t_{\rm f})\Phi_{t_{\rm f}}[\rho_{\rm i}]) \\
 &- 2\,{\rm Tr}(\Phi_{t_{\rm f}}[\rho_{\rm i}]{H}(t_{\rm f}))\,{\rm Tr}(\rho_{\rm i}{H}(t_{\rm i})),
\end{aligned}
\end{equation}
which coincides with what is achieved through the MLL
protocol only for initial pure states and through the TPM protocol
only if the initial state is an eigenstate of ${H}(t_{\rm i})$.
Eq.~\eqref{second_mom} can be cast in a form that lets 
the contribution of the initial coherence emerge clearly as
\begin{equation}\label{division}
\begin{aligned}
 \langle\Delta E^2\rangle &=\,\langle\Delta E^2\rangle_{\mathcal{P}}+{\rm Tr}({H}^2(t_{\rm f})\Phi_{t_{\rm f}}[\chi])\\ 
 &- 2\,{\rm Tr}(\Phi_{t_{\rm f}}[\chi]{H}(t_{\rm f}))\,{\rm Tr}(\mathcal{P}{H}(t_{\rm i})),
\end{aligned} 
\end{equation}
where $\langle\Delta E^2\rangle_{\mathcal{P}}$ is obtained
from Eq.~\eqref{second_mom} by replacing $\rho_{\rm i}\rightarrow \mathcal{P}$.
It should be noted that, if the initial state $\rho_{\rm i}$
is such that $\mathcal{P}$ is a projector, then
$\langle\Delta E^2\rangle_{\mathcal{P}}=\langle \Delta E^2\rangle_{\rm TPM}$
and all the differences in the second moments are originated
by coherence terms in $\rho_{\rm i}$. The latter, indeed,
are unavoidably destroyed by applying the TPM protocol.

\noindent
    {\bf Characteristic function and physical meaning.--}The information about
    the statistical moments of the distribution of energy changes
    is encoded in the characteristic function
    $\mathcal{G}(u) \equiv \langle e^{iu\Delta E}\rangle_{{\rm P}_{\rm coh}} =
    \int d\Delta E\,e^{iu\Delta E}{\rm P}_{\rm coh}(\Delta E)$, $u\in\mathbb{C}$
    corresponding to the probability distribution
    ${\rm P}_{\rm coh}(\Delta E)$. As the outcomes $\{E_{\rm f}^{(k)}\}$
    of the final energy  measurement are statistically independent
    from the initial virtual ones $\{E_{\rm i}^{(\ell)}\}$, we have
\begin{equation}\label{G_u}
    \mathcal{G}(u) = {\rm Tr}(e^{-iu\mathcal{H}(t_{\rm i})}\rho_{\rm i})\,{\rm Tr}(e^{iu\mathcal{H}(t_{\rm f})}\Phi_{t_{\rm f}}[\rho_{\rm i}]),
\end{equation}
from which we see that the fluctuations of $\Delta E$ originate
both from the action of the dynamical map $\Phi_t[\rho]$ on the initial state
of the system and the uncertainty in its energy at $t=t_{\rm i}$.

Let us now show how such generating function leads naturally
to a statement highlighting the deviation of the EPM-inferred statistics
from a standard fluctuation theorem~\cite{CampisiRMP2011,EspositoRMP2009}.
To this goal, we consider the logarithm of $\ln\mathcal{G}(i\beta)$,
where $\beta$ is a \textit{reference} inverse temperature
(to be taken as a free parameter) and introduce the equilibrium
reference states
$\rho_{\rm i(f)}^{\rm th} \equiv e^{-\beta {H}({ t}_{\rm i(f)})}/Z_{\rm i(f)}$ with
$Z_{\rm i(f)} \equiv {\rm Tr}(e^{-\beta {H}({\rm t}_{\rm i(f)})})$
the corresponding partition functions. Assuming the initial state
$\rho_{\rm i} = \rho_{\rm i}^{\rm th}+\chi$ and a unital dynamical
map~\footnote{The unitality of the map, while allowing a direct comparison
  with the standard Jarzynski equality
  $\ln\mathcal{G}_{\rm TPM}(i\beta) = -\beta\Delta F$, can be relaxed.
  We refer to the appendix for further details.} we get
\begin{equation}\label{eq:Jarzynski_like_rel}
    \langle e^{-\beta(\Delta E{-}\Delta F)}\rangle{=}d\left({\rm Tr}\left(\rho_{\rm f}^{\rm th}\,\Phi_{t_{\rm f}}[\rho_{\rm i}^{\rm th}]\right){+}{\rm Tr}\left(\rho_{\rm f}^{\rm th}\,\Phi_{t_{\rm f}}[\chi]\right)\right),
\end{equation}
where $\Delta F$ is the free energy difference
(details on the derivation of this result are reported in the appendix.
The right-hand-side of Eq.~\eqref{eq:Jarzynski_like_rel} deviates from unity,
i.e., from a standard fluctuation theorem, in light of two distinct factors.
The first,
$d\,{\rm Tr}(\rho_{\rm f}^{\rm th}\,\Phi_{t_{\rm f}}[\rho_{\rm i}^{\rm th}])$,
is the additional uncertainty introduced
by not performing the initial energy measurement. This extra uncertainty
is present even for $\chi=0$. The second term
$d\,{\rm Tr}(\rho_{\rm f}^{\rm th}\,\Phi_{t_{\rm f}}[\chi])$,
quantifies the deviation due to the initial quantum coherences
alone and thus bridges stochastic thermodynamics and genuine
quantum signatures of open system dynamics. Eq.~\eqref{eq:Jarzynski_like_rel}
is one of the main results of this paper.

\begin{figure}[t!]
\centering
\includegraphics[scale=0.45]{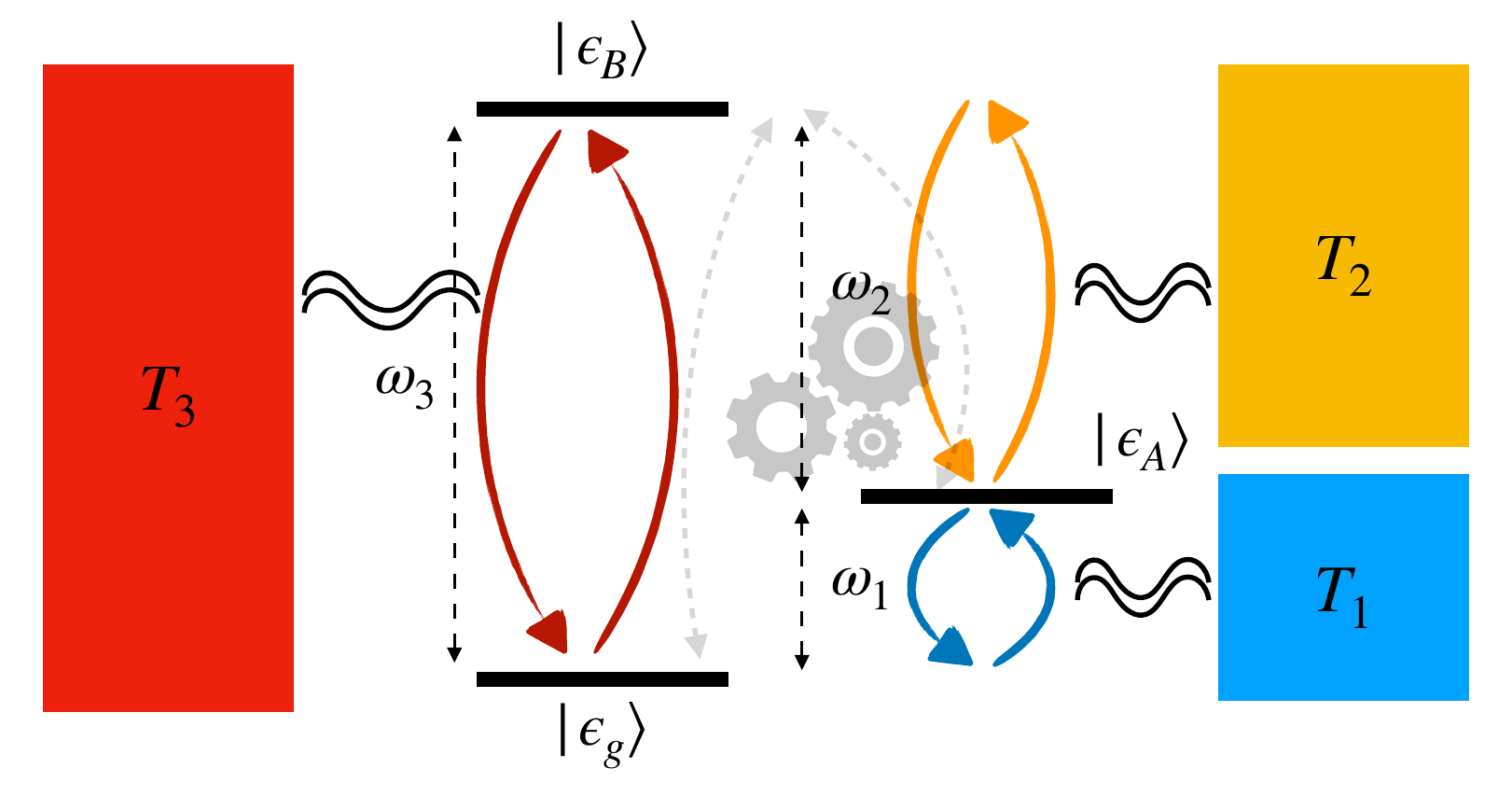}
\caption{Pictorial illustration of a $3$-level system coupled to
  three thermal baths at different temperatures $T_k$, $k=1,2,3$,
  and externally driven by a time-dependent Hamiltonian term (in light-grey)
  with $\omega_3=\omega_1+\omega_2$.}
\label{fig3level}
\end{figure}

\noindent
{\bf Numerical example.--}
In order to illustrate the effect of initial coherence on energy
fluctuations as singled out by our EPM protocol, we address a simple yet
physically relevant example.  
Let us consider a three-level quantum system in interaction with three
thermal reservoirs and driven by a time-dependent Hamiltonian
[cf. Fig.\,\ref{fig3level}]. The open-system dynamics is described by
the Lindblad master equation 
\begin{equation}\label{me3main}
  \dot{\rho}_t=-i[H+H_{\rm{drive}},\rho]+\sum_{i\neq j=1}^{3}
  \left( L_{ij}\rho L_{ij}^\dag-\frac{1}{2}\{L_{ij}^\dag L_{ij},\rho\}\right).
\end{equation}
Here, $L_{ij}\equiv\sqrt{\eta_{ij}}|\epsilon_i\rangle\!\langle \epsilon_j|$
is an environment-induced jump operator acting on the system at rate
$\eta_{ij}$ (see appendix). The free Hamiltonian of the system is
$H=\omega_3|\epsilon_B\rangle\!\langle \epsilon_B|+\omega_1 |\epsilon_A\rangle\!\langle \epsilon_A|$,
where $\{\ket{\epsilon_g},\ket{\epsilon_A},\ket{\epsilon_B}\}$
are the three levels of the system with associated energies
$0,\omega_1$ and $\omega_3$, respectively. The driving term is chosen
as $H_{\rm{drive}}=(g(t)\ket{\epsilon_g}+f(t)\ket{\epsilon_A})\bra{\epsilon_B}+{\rm h.c.}$, with $f(t)$ and $g(t)$ time-dependent coupling rates. 

This setting has been used as an archetypal quantum autonomous thermal
machines and studied in a variety of different
configurations~\cite{ScovilPRL59,PalaoPRE2001,KosloffARPC2014}.

\begin{figure*}[ht!]
{\bf (a)}\hskip8cm{\bf (b)}\\
\centering
\includegraphics[scale=0.58]{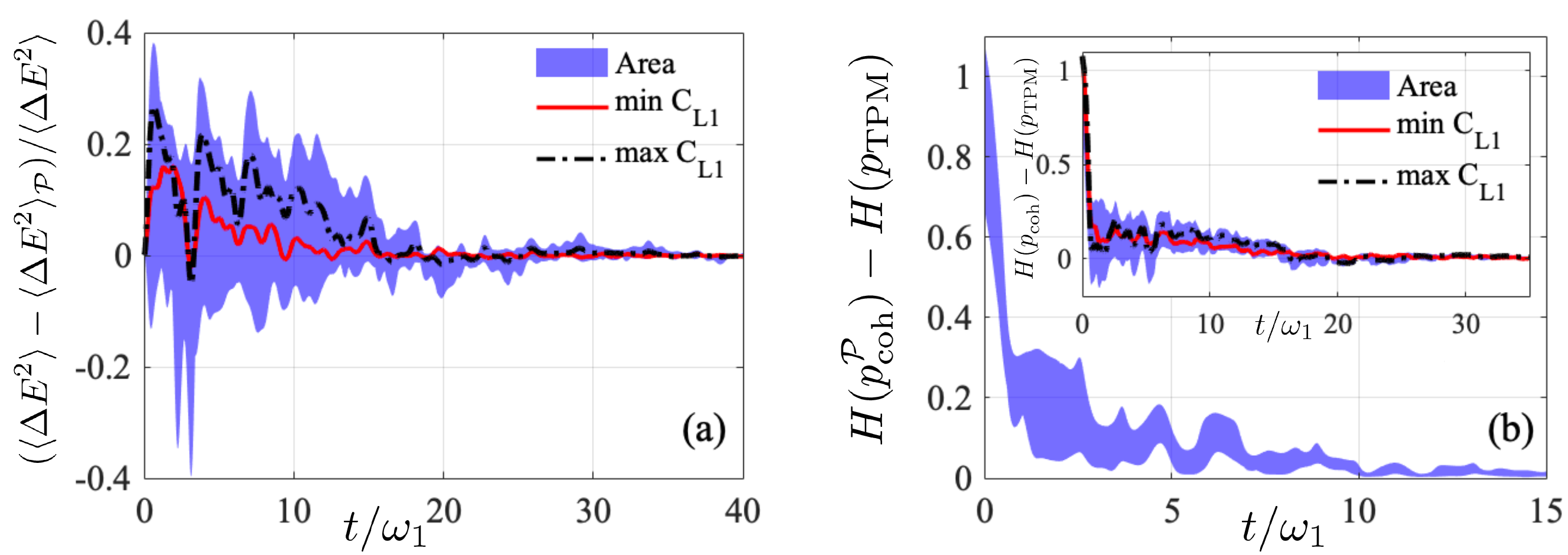}
\caption{\textbf{(a)}: $\left(\langle\Delta E^2\rangle -
  \langle\Delta E^2\rangle_{\mathcal{P}}\right)/\langle\Delta E^2\rangle$
  as a function of time for the driven $3$-level system
  with dynamics as given in Eq.~\eqref{me3main}. \textbf{(b)}:
  Shannon entropy difference between the EPM scheme and TPM one
  with no coherence in the energy basis.
  Inset: Difference between the Shannon entropies for the full,
  non-diagonal, density matrices $\rho_{\rm i}$. In both figures,
  the blue-shaded regions encompass the values obtained by numerically
  evaluating the statistics of $\Delta E$ for $10^3$ random initial states
  --- uniformly sampled by respecting the Haar measure of the space of
  $3\times 3$ density operators. The red solid lines
  (black dash-dotted line) denote the corresponding curves obtained
  by taking as initial $\rho_{\rm i}$
  the quantum state in such sample with the lowest
  (highest) value of coherence according to the measure of quantum
  coherence $C_{L1}$~\cite{Baumgratz} (notice that consequently
  the red and black dash-dotted line lines are not plotted in the
  main right figure). 
  The parameters used in the simulations are
  $\omega_k = k\omega_1$, $\gamma=0.1\omega_1$, $\beta_1=3$,
  $\beta_2=1$, $\beta_3=2$, $g(t)=1.5\sin^2(t)$ and $f(t)=1.5[1-\sin^2(2t)]$
  [with  $\omega_1 = \hbar = k_B = 1$].}
\label{fig:2nd_moment_coh}
\end{figure*}

When the three reservoirs have the same temperature, and in the
absence of an external driving, the system relaxes to the
equilibrium state
$\rho^\text{th}_{\infty} \equiv e^{-\beta H}/{\rm Tr}(e^{-\beta H})$.
In this simple scenario, independently of the initial state,
the distribution in Eq.~\eqref{prob_distr} converges in the asymptotic
limit to the one resulting from the MLL and TPM schemes. The same is true,
in general, for any dynamical map allowing for a unique fixed-point.
Differently -- and in accordance with our general discussion --
at intermediate times the statistics from the three approaches differ
even in the absence of coherence in the initial state.

To illustrate our findings, in Fig.~\ref{fig:2nd_moment_coh}\,(a)
we plot the temporal behavior of
$1 - \langle\Delta E^2\rangle_{\mathcal{P}}/\langle\Delta E^2\rangle$,
evaluated using Eq.~\eqref{division}. This is exactly the contribution
to the second moment of $\Delta E$ originated by the
initial-coherence terms in $\rho_{\rm i}$, which our protocol allows
us to neatly identify. In this specific case, we observe that
the contribution of the initial coherence can be as large as $40\%$
of the total value taken by the second moment of the energy fluctuations.
In Fig.~\ref{fig:2nd_moment_coh}\,(b), we show the discrepancies between
the Shannon entropy of the EPM-based energy-change probability distribution
in the absence of initial coherence and that stemming from
TPM-based predictions. As previously discussed,
this difference quantifies the extra uncertainty, with respect to the
TPM scheme, due to \textit{not} performing the initial
energy measurement. The inset shows how coherences in the
initial state can make the entropy difference negative.
This implies that initial coherence could compensate
for the extra-uncertainty due to the virtual initial measurement,
thus providing a statistically more informative characterisation of
energy fluctuations.

From these results, we deduce that the quantum coherence initially present
in $\rho_{\rm i}$ (in the system energy basis) has an active role
in the first part of the system evolution and is propagated
thanks to the action of the driving Hamiltonian. This phenomenon
is well-captured by the energy-change fluctuations quantified by
the EPM protocol. In this specific example, the contribution of
the coherence is suppressed at long times. This is due to the fact
that the dynamics reaches a (time-dependent) fixed-point,
independently of the initial state. Consistently with our previous
discussion, in this scenario the EPM probability distribution converges
to the TPM one. It should also be noted that, while the initial
coherence has a relevant impact on the statistics of the energy fluctuations,
the time behavior of
$\langle\Delta E^2\rangle - \langle\Delta E^2\rangle_{\mathcal{P}}$,
which is the term related to the initial coherence, is never
monotonic with the amount of such  coherences.
We show this in Fig.~\ref{fig:2nd_moment_coh} using the measure of
quantum coherence $C_{L1} \equiv \frac{1}{2}\sum_{i,j,\,i\neq j}|\rho_{ij}|$
put forward in Ref.~\cite{Baumgratz}:
the curves corresponding to initial states with maximum coherence
never maximize the difference between the results from EPM and
TPM distributions, even though they are rather close to it.

\noindent
{\bf Conclusions.--}
We have introduced a novel {operational} protocol for the evaluation
of the energy-change fluctuations resulting from general open
quantum-system dynamics. Our EPM protocol is able to suitably
take into account the presence of quantum coherence
in the initial state of the system without requiring information
on the system dynamics, which casts it apart from other schemes
such as~\cite{DeffnerPRE2016,Sone2020}. Moreover, it does
not need the initial preparation of the system in an eigenstate
of its density operator, thus making it different from
the scheme put forward recently in Ref.~\cite{MicadeiPRL2020}. 

Besides the knowledge of the initial state,
our EPM protocol solely relies on the final energy measurement.
The scheme allows to neatly single out the contribution of
the initial coherence to the energy fluctuation statistics.
These contributions are, in general, not negligible and
significantly impact the energy-change statistics. 

The EPM approach could be more conducive of experimental
validation than the notoriously challenging TPM one, and thus
holds the potential to enlarge the range of systems whose energy-change
fluctuations could be tested. For instance, it can significantly
help in the case of systems with highly degenerate energy levels,
as it occurs in many-body physics. For an initial state involving
only levels within such degenerate subspace and a dynamics that leaves
the latter invariant, the TPM scheme would return vanishing energy
fluctuations. In contrast, our EPM protocol would allow for
the characterization of the energy-change statistics resulting
from the initial coherence alone, thus showcasing its
sensitivity to the coherence features of quantum systems.

\paragraph*{Acknowledgments.--}

The authors gratefully acknowledge
N. Fabbri, M. Garc\'ia D\'iaz, G. Guarnieri, G.T. Landi, M. Lostaglio, S. Hern\'andez-G\'omez and F. Poggiali
for fruitful discussions and comments. This work was supported by MISTI
Global Seed Funds MIT-FVG Collaboration Grant
"NV centers for the test of the Quantum Jarzynski Equality (NVQJE)",
H2020-FETOPEN-2018-2020 project PATHOS (grant nr.\,828946),
UNIFI grant Q-CODYCES, the MSCA IF project pERFEcTO (grant nr. 795782),
the H2020-FETOPEN-2018-2020 project TEQ (grant nr.~766900),
the DfE-SFI Investigator Programme (grant 15/IA/2864),
COST Action CA15220, the Royal Society Wolfson Research Fellowship
(RSWF\textbackslash R3\textbackslash183013),
the Royal Society International Exchanges Programme
(IEC\textbackslash R2\textbackslash192220),
the Leverhulme Trust Research Project Grant (grant nr.~RGP-2018-266) and
the CNR/RS (London) project
``Testing fundamental theories with ultracold atoms" 


\begin{widetext}

\section{Appendix}

\setcounter{equation}{0}
\setcounter{figure}{0}
\setcounter{table}{0}
\renewcommand{\theequation}{S\arabic{equation}}
\renewcommand{\thefigure}{S\arabic{figure}}

\section{Classical uncertainty on the initial state}

The operational protocol that we are introducing in this paper does not reproduce
the same results of the  two-point measurement (TPM) scheme even in the absence
of coherence in the initial state $\rho_{\rm i}$. There is indeed a discrepancy originating
from a classical uncertainty on even diagonal (in the initial energy basis)
$\rho_{\rm i}$ that is retained in our scheme. Despite this aspect
is in agreement with the theses of the no-go theorem~\cite{Perarnau-LlobetPRL2017}
as explained in the main text, it is worth understanding it in more detail. In this regard,
let us now substitute the density operator
$\varrho \equiv \sum_{r}p_{\rm i}^{(r)}\rho_{\rm i}^{(r)} = \sum_{r}p_{\rm i}^{(r)}\Pi_{\rm i}^{(s)}$
(mixed quantum state diagonal in the energy basis of the system at
$t_{\rm i}$, i.e., $[\varrho,\mathcal{H}(t_{\rm i})]=0$) as input quantum state
$\rho_{\rm i}$ in Eq.~\eqref{prob_distr} of the main text. One finds that
\begin{eqnarray}\label{eq:eq_1}
{\rm P}_{\rm coh}(\Delta E) &=& \sum_{k,\ell}p_{\rm i}^{(\ell)}p_{\rm f}^{(k)}\delta(\Delta E - \Delta E_{k,\ell})
= \sum_{k,\ell}{\rm Tr}(\Pi_{\rm i}^{(\ell)}\rho_{\rm i}){\rm Tr}(\Pi_{\rm f}^{(k)}\Phi_{t_{\rm f}}[\rho_{\rm i}])\delta(\Delta E - \Delta E_{k,\ell})\nonumber \\
&=& \sum_{k,\ell,r_{1},r_{2}}p_{\rm i}^{(r_1)}p_{\rm i}^{(r_2)}{\rm Tr}(\Pi_{\rm i}^{(\ell)}\Pi_{\rm i}^{(r_1)}){\rm Tr}(\Pi_{\rm f}^{(k)}\Phi_{t_{\rm f}}[\Pi_{\rm i}^{(r_2)}])\delta(\Delta E - \Delta E_{k,\ell})\nonumber \\
&=& \sum_{k,r_{1},r_{2}}p_{\rm i}^{(r_1)}p_{\rm i}^{(r_2)}{\rm Tr}(\Pi_{\rm f}^{(k)}\Phi_{t_{\rm f}}[\Pi_{\rm i}^{(r_2)}])\delta(\Delta E - \Delta E_{k,r_{1}})
= \sum_{k,r_{1},r_{2}}p_{\rm i}^{(r_{1})}p_{\rm f,i}^{(k,r_{2})}\delta(\Delta E - \Delta E_{k,r_{1}}) \ ,
\end{eqnarray}
where we have used the relations
${\rm Tr}(\Pi_{\rm i}^{(\ell)}\Pi_{\rm i}^{(r_1)}) = \delta(\ell-r_{1})$ and
$p_{\rm f,i}^{(k,r_{2})} \equiv p_{\rm i}^{(r_2)}{\rm Tr}(\Pi_{\rm f}^{(k)}\Phi_{t_{\rm f}}[\Pi_{\rm i}^{(r_2)}]) = p_{\rm i}^{(r_2)}p_{\rm f|i}^{(k,r_2)}$
with $p_{\rm f,i}^{(k,r_{2})}$ joint probabilities.

From Eq.~(\ref{eq:eq_1}) one can deduce that
${\rm P}_{\rm coh}(\Delta E) = {\rm P}_{\rm TPM}(\Delta E)$
if and only if the initial state
is chosen as one of the eigenstates of the initial Hamiltonian,
such that $\Delta E_{k,r_{1}} = \Delta E_{k,r_{2}}$. Indeed, in such a case
\begin{equation}\label{eq:eq_2}
{\rm P}_{\rm coh}(\Delta E) = \sum_{r_1}p_{\rm i}^{(r_1)}\sum_{k,r_{2}}p_{\rm f,i}^{(k,r_{2})}\delta(\Delta E - \Delta E_{k,r_{2}})
= \sum_{r_1}p_{\rm i}^{(r_1)}{\rm P}_{\rm TPM}(\Delta E) = {\rm P}_{\rm TPM}(\Delta E) \ .
\end{equation}
It is then clear that an initial uncertainty on which eigenstate of the Hamiltonian
needs to be propagated, due to the fact that in our protocol the initial measurement
is \textit{virtual}, determines an additional uncertainty on the energy statistics,
which is reflected in the discrepancy between the two methods. The latter is here
provided by the arbitrariness of the possible inequality $\Delta E_{k,r_{1}} \neq \Delta E_{k,r_{2}}$.

\section{Recovering the TPM statistics}
\label{sec:app_rec_class_limit}
As stated before, the energy change probability distribution ${\rm P}_{\rm coh}$ does not reduce
to the one from the TPM scheme unless the initial state of both protocol is an energy
eigenstate. Considering again an initial state diagonal in the energy eigenbasis,
it is easy to see from that, in order to find the same statistics of $\Delta E$
as given by a TPM protocol, Eq.~(\ref{prob_distr}) has to be used as many times
as the number of probabilities $p_{\rm i}^{(r)}$ defining the initial density operator
$\varrho$, initializing each time the quantum system in one of the projectors $\Pi_{\rm i}^{(r)}$.
In doing this, the corresponding probability distribution of $\Delta E$ turns out to be
\begin{eqnarray}\label{conventional_prob_E}
  &{\rm P}_{\rm coh}(\Delta E) = \displaystyle{\sum_{r}p_{\rm i}^{(r)}\sum_{k,\ell}\delta(\Delta E - \Delta E_{k,\ell})\delta(\ell-r){\rm Tr}(\Phi_{t_{\rm f}}[\Pi_{\rm i}^{(r)}]\Pi_{\rm f}^{(k)})}&\nonumber \\
  &= \displaystyle{\sum_{k,r}\delta(\Delta E - \Delta E_{k,r})p_{\rm f|i}^{(k,r)}p_{\rm i}^{(r)}} \equiv {\rm P}_{\rm TPM}(\Delta E) \,&
\end{eqnarray}
where $p_{\rm f|i}^{(k,r)} \equiv {\rm Tr}(\Phi_{t_{\rm f}}[\Pi_{\rm i}^{(r)}]\Pi_{\rm f}^{(k)})$
is the transition probability to measure the final energy $E^{(k)}_{\rm f}$ conditioned to have
obtained $E^{(r)}_{\rm i}$ at $t=t_{\rm i}$. Only in this way, the proposed formalism falls
into the category of FT protocols~\cite{LostaglioPRL2018}, so that we can recover
the conventional statistics of energy change as provided by the TPM scheme.
This result is not surprising, since we are now analyzing a situation in which a
possible first energy measurement at $t=t_{\rm i}$ would
not introduce any disturbance to the evolution of the system. As a further remark,
also notice that with this approach 
the notion of quasi-probabilities
is not directly used~\cite{AllahverdyanPRE2014,LevyArxiv2019}.

\section{Analysis of the $1$st and $2$nd energy statistical moments}
In this section, we provide the analytical expressions of the $1$st and $2$nd
statistical moments of the proposed energy change distribution in comparison
with the ones obtained by the TPM protocol and the Micadei-Landi-Lutz (MLL)
protocol~\cite{MicadeiPRL2020}. In doing this, we recall that the initial state $\rho_{\rm i}$
is expressed in Ref.~\cite{MicadeiPRL2020} in terms of its eigenstates
with notation $\sum_{s}p^{(s)}|s\rangle\!\langle s|$, which is the same that we will
use in the following.
We list below all the formulas of the joint probability $p(E_{\rm i}^{(\ell)},E_{\rm f}^{(k)})$
and the $1$st and $2$nd statistical moments of $\Delta E$ that one can obtain from the three methods.\\ \\
\textbf{EPM protocol proposed in the present paper:}
\begin{eqnarray}
    && p(E_{\rm i}^{(\ell)},E_{\rm f}^{(k)}) = {\rm Tr}(\rho_{\rm i}\Pi_{\rm i}^{(\ell)})\,{\rm Tr}(\Phi_{t_{\rm f}}[\rho_{\rm i}]\Pi_{\rm f}^{(k)}) \\
    && \langle\Delta E\rangle = {\rm Tr}(\mathcal{H}(t_{\rm f})\Phi_{t_{\rm f}}[\rho_{\rm i}]) - {\rm Tr}(\mathcal{H}(t_{\rm i})\rho_{\rm i}) \\
    && \langle\Delta E^2\rangle = {\rm Tr}(\mathcal{H}^2(t_{\rm i})\rho_{\rm i}) + {\rm Tr}(\mathcal{H}^2(t_{\rm f})\Phi_{t_{\rm f}}[\rho_{\rm i}]) - 2\,{\rm Tr}(\Phi_{t_{\rm f}}[\rho_{\rm i}]\mathcal{H}(t_{\rm f}))\,{\rm Tr}(\rho_{\rm i}\mathcal{H}(t_{\rm i})) \ . \label{eq:app_2nd_moment}
\end{eqnarray}\\
\textbf{Micadei-Landi-Lutz protocol:}
\begin{eqnarray}
    && p(E_{\rm i}^{(\ell)},E_{\rm f}^{(k)}) = \sum_s p^{(s)}{\rm Tr}(|s\rangle\!\langle s|\Pi_{\rm i}^{(\ell)})\,{\rm Tr}(\Phi_{t_{\rm f}}[|s\rangle\!\langle s|]\Pi_{\rm f}^{(k)}) \\
    && \langle\Delta E\rangle = {\rm Tr}(\mathcal{H}(t_{\rm f})\Phi_{t_{\rm f}}[\rho_{\rm i}]) - {\rm Tr}(\mathcal{H}(t_{\rm i})\rho_{\rm i}) \\
    &&  \langle\Delta E^2\rangle = {\rm Tr}(\mathcal{H}^2(t_{\rm i})\rho_{\rm i}) + {\rm Tr}(\mathcal{H}^2(t_{\rm f})\Phi_{t_{\rm f}}[\rho_{\rm i}]) - 2\,\sum_s p^{(s)}{\rm Tr}(\Phi_{t_{\rm f}}[|s\rangle\!\langle s|]\mathcal{H}(t_{\rm f}))\,{\rm Tr}(|s\rangle\!\langle s|\mathcal{H}(t_{\rm i})) \ . \label{eq:app_2nd_moment_MLL}
\end{eqnarray}\\
\textbf{TPM protocol:}
\begin{eqnarray}
    && p(E_{\rm i}^{(\ell)},E_{\rm f}^{(k)}) = {\rm Tr}(\rho_{\rm i}\Pi_{\rm i}^{(\ell)})\,{\rm Tr}(\Phi_{t_{\rm f}}[\Pi_{\rm i}^{(\ell)}]\Pi_{\rm f}^{(k)}) \\
    && \langle\Delta E\rangle = {\rm Tr}\left(\mathcal{H}(t_{\rm f})\Phi_{t_{\rm f}}\left[\sum_{\ell}{\rm Tr}(\rho_{\rm i}\Pi_{\rm i}^{(\ell)})\Pi_{\rm i}^{(\ell)}\right]\right) - {\rm Tr}(\mathcal{H}(t_{\rm i})\rho_{\rm i}) \\
    &&  \langle\Delta E^2\rangle = {\rm Tr}(\mathcal{H}^2(t_{\rm i})\rho_{\rm i}) + 
    {\rm Tr}\left(\mathcal{H}^2(t_{\rm f})\Phi_{t_{\rm f}}\left[\sum_{\ell}{\rm Tr}(\rho_{\rm i}\Pi_{\rm i}^{(\ell)})\Pi_{\rm i}^{(\ell)}\right]\right) - 2\,\sum_{\ell}E_{\rm i}^{(\ell)}{\rm Tr}(\mathcal{H}(t_{\rm f})\Phi_{t_{\rm f}}[\Pi_{\rm i}^{(\ell)}])\,{\rm Tr}(\rho_{\rm i}\Pi_{\rm i}^{(\ell)}) \ .
\end{eqnarray}

Within the TPM protocol, an initial measurement of the system Hamiltonian at $t=t_{\rm i}$
and a final one at $t=t_{\rm f}$ on the conditional evolved states are performed.
In order to get the corresponding conditional probability, the system has to be
separately initialized in each eigenstate of $\mathcal{H}(t_{\rm i})$, respectively.

Concerning the MLL protocol, for the sake of experimentally characterise the energy change
probability distribution, one needs to initialize the system in the eigenstates
of the initial state $\rho_{\rm i}$. This operation could be equivalently carried
on by performing an initial measurement of the observable
$\mathcal{O}\equiv\sum_{s}o_{s}|s\rangle\!\langle s|$, in general not
commuting with $\mathcal{H}(t_{\rm i})$. Indeed, according to the MLL protocol,
the final energy measurement is performed on the evolved eigenstate
($|s\rangle\!\langle s|$) of the initial state and the results are then
weighted with the probabilities $\{p^{(s)}\}$.

Finally, in our EPM protocol, the initial state $\rho_{\rm i}$ is arbitrary
and the final energy measurement is performed on the evolved initial state
without any need to initialize the system in a different state.
The energy change probability distribution is obtained by weighting these
final probabilities with the ones concerning the initial virtual energy measurement,
which are accessible from the knowledge of the initial state. 

One can observe that the average energy change $\langle\Delta E\rangle$
provided by our protocol and the MLL protocol are the same, differently to
the one from the TPM protocol for which the mean final energy measured at $t=t_{\rm f}$
does not contain any contributions from initial coherence terms in $\rho_{\rm i}$.
Furthermore, regarding the $2$nd moment $\langle\Delta E^{2}\rangle$, the three protocols
differ again for the way in which the initial energy outcomes
(eigenvalues of $\mathcal{H}(t_{\rm i})$) are taken into account in
relation to $\rho_{\rm i}$. Only our (operational) method makes no assumptions
about $\rho_{\rm i}$, since we completely remove the need to perform
any initial projective measurement. 
However, in general, if the initial state $\rho_{\rm i}$ is pure,
the second moments~\eqref{eq:app_2nd_moment} and \eqref{eq:app_2nd_moment_MLL} coincide,
while, as shown above in 
Section II, $\eqref{eq:app_2nd_moment}$ coincides with the second moment
obtained by applying the TPM protocol for an initial state corresponding to an
eigenstate of $\mathcal{H}(t_{\rm i})$.

It is also interesting to note that the probability distribution from our protocol
corresponds to the product of the marginals of the MLL-protocol probability distribution~\footnote{The authors thank Gabriel T. Landi for pointing this out to us.}.
In particular, the (informational) price that we have to pay due to \textit{not}
performing any initial measurement, with respect to the MLL protocol that requires
a greater knowledge of the state and dynamics, can be quantified by the mutual
information between the two probability distributions, i.e.,
\begin{equation}
    \mathcal{I}({\rm P}_{\rm MLL},{\rm P}_{\rm coh})=\sum_{k,\ell}p_{\rm MLL}^{(k,\ell)}\log\left(p_{\rm MLL}^{(k,\ell)}/p_{\rm coh}^{(k,\ell)}\right).
\end{equation}
$\mathcal{I}({\rm P}_{\rm MLL},{\rm P}_{\rm coh})$ encodes the cost
of our assumption of the statistical independence between the final energy measurement
and the initial virtual one with respect to the MLL scheme.
\\ \\
Let us summarize what we have discussed so far concerning the connection of the proposed protocol with the TPM and MLL schemes:
\begin{itemize}
    \item 
    For an initial state $\rho_{\rm i}$ diagonal in the energy eigenbasis, the MLL and TPM protocols provide the same joint probability $p(E_{\rm i}^{(\ell)},E_{\rm f}^{(k)})$, while our protocol differ from them.
    \item 
    For an initial pure state, not necessarily an eigenstate of the initial Hamiltonian, the joint probabilities from our method and the MLL protocol coincide.
    \item 
    In the special case of initial pure energy eigenstate, all three protocols give the same result.
\end{itemize}
A first element of difference between our protocol and the TPM and MLL ones
is given by a classical uncertainty on the initial state $\rho_{\rm i}$.
This is due to the fact that we are assuming to not know the single pure
components that decompose the initial state $\rho_{\rm i}$, or at least the effect
of the dynamics on them separately. Operationally, both the TPM (explicitly) and the
MLL (implicitly) need to assume the knowledge about the evolution of the pure components
of the system initial state (either in the energy eigenbasis or in its eigenbasis),
which are then evolved and give rise to conditional probabilities. The proposed protocol
does not assume this knowledge and it is thus nicely amenable
for experimental implementations with minimal resources.

\section{Energy change characteristic function}
Here, we provide the mathematical details for the derivation of the characteristic
function associated to the energy change distribution both from the proposed method and the
MLL and TPM protocols. 
The characteristic function from the three methods are respectively equal to 
\begin{align}
    & \mathcal{G}(u) = {\rm Tr}(e^{-iu\mathcal{H}(t_{\rm i})}\rho_{\rm i})\,{\rm Tr}(e^{iu\mathcal{H}(t_{\rm f})}\Phi_{t_{\rm f}}[\rho_{\rm i}])\label{eq:app_G_coh} \\
    & \mathcal{G}_{\rm MLL}(u) = \sum_s p^{(s)} {\rm Tr}\left(|s\rangle\!\langle s| e^{-iu\mathcal{H}(t_{\rm i})}\right)\,{\rm Tr}\left(\Phi_{t_{\rm f}}[|s\rangle\!\langle s|]e^{iu\mathcal{H}(t_{\rm f})}\right) \\
    & \mathcal{G}_{\rm TPM}(u) = {\rm Tr}(e^{iu\mathcal{H}(t_{\rm f})}\Phi_{t_{\rm f}}[e^{-iu\mathcal{H}(t_{\rm i})}\rho_{\rm i}]) 
\end{align}
with $u\in\mathbb{C}$ complex number. Also at the level of the characteristic function of
the energy change distribution, we can single out coherence contributions. In particular,
by taking $\rho_i=\mathcal{P}+\chi$ in~\eqref{eq:app_G_coh}, one has
\begin{align}
    \mathcal{G}(u) &= {\rm Tr}(e^{-iu\mathcal{H}(t_{\rm i})}\rho_{\rm i})\,{\rm Tr}(e^{iu\mathcal{H}(t_{\rm f})}\Phi_{t_{\rm f}}[\rho_{\rm i}])\nonumber \\
    &={\rm Tr}(e^{-iu\mathcal{H}(t_{\rm i})}\mathcal{P})\,{\rm Tr}(e^{iu\mathcal{H}(t_{\rm f})}\Phi_{t_{\rm f}}[\mathcal{P}])+{\rm Tr}(e^{-iu\mathcal{H}(t_{\rm i})}\mathcal{P})\,{\rm Tr}(e^{iu\mathcal{H}(t_{\rm f})}\Phi_{t_{\rm f}}[\chi])\nonumber \\
    &\equiv\mathcal{G}_{\mathcal{P}}(u)+\mathcal{G}_{\mathcal{\chi}}(u), 
\end{align}
where $\mathcal{G}_{\mathcal{P}}(u) \equiv {\rm Tr}(e^{-iu\mathcal{H}(t_{\rm i})}\mathcal{P})\,{\rm Tr}(e^{iu\mathcal{H}(t_{\rm f})}\Phi_{t_{\rm f}}[\mathcal{P}])$ and
$\mathcal{G}_{\mathcal{\chi}}(u) \equiv {\rm Tr}(e^{-iu\mathcal{H}(t_{\rm i})}\mathcal{P})\,{\rm Tr}(e^{iu\mathcal{H}(t_{\rm f})}\Phi_{t_{\rm f}}[\chi])$. As a result,
$\mathcal{G}_{\chi}=0$ when $\chi=0$ and $\mathcal{G}_{\mathcal{P}}=\mathcal{G}_{\rm TPM}$
as far as $\mathcal{P}$ is a projector associated to a system energy eigenspace. 
Now, given the expressions of $\mathcal{G}$ and $\mathcal{G}_{\rm TPM}$,
we derive their logarithm. In this way, as explained in the main text,
we are able to provide some physical interpretations of our findings. In doing this,
let us introduce the inverse temperature $\beta$ taken as a free reference parameter
and the two thermal (reference) states 
\begin{equation}\label{eq:app_ref_th_states}
\rho_{\rm i(f)}^{\rm th} \equiv \displaystyle{\frac{e^{-\beta\mathcal{H}(t_{\rm i(f)})}}{Z_{\rm i(f)}}}
\end{equation}
referring, respectively, to the initial and final time instants of the protocols.
In Eq.\,(\ref{eq:app_ref_th_states}),
$Z_{\rm i(f)} \equiv {\rm Tr}(e^{-\beta\mathcal{H}(t_{\rm i(f)})})$, such that
the free-energy difference $\Delta F$ in the time interval $[t_{\rm i},t_{\rm f}]$ is equal,
as usual, to  
\begin{equation}
\Delta F = -\beta^{-1}\ln\left(\frac{Z_{\rm f}}{Z_{\rm i}}\right) \ . 
\end{equation}
The logarithms of $\mathcal{G}$ and $\mathcal{G}_{\rm TPM}$, computed at $u=i\beta$,
are provided by the following relations
\begin{eqnarray}
\ln \mathcal{G}(i\beta) &=& \ln{\rm Tr}(e^{\beta\mathcal{H}(t_{\rm i})}\rho_{\rm i})+\ln\{Z_{\rm f}{\rm Tr}(\rho^{\rm th}_{\rm f}\Phi_{t_{\rm f}}[\rho_{\rm i}])\}
= \ln Z_{\rm f} + \ln\{Z_{\rm i}^{-1}{\rm Tr}(Z_{\rm i}\,e^{\beta\mathcal{H}(t_{\rm i})}\rho_{\rm i})\} + \ln{\rm Tr}(\rho^{\rm th}_{\rm f}\Phi_{t_{\rm f}}[\rho_{\rm i}])\nonumber \\
&=& -\beta\Delta F + \ln{\rm Tr}((\rho_{\rm i}^{\rm th})^{-1}\rho_{\rm i}) + \ln{\rm Tr}(\rho_{\rm f}^{\rm th}\Phi_{t_{\rm f}}[\rho_{\rm i}]),
\end{eqnarray}
and
\begin{eqnarray}
\ln\mathcal{G}_{\rm TPM}(i\beta)&=&\ln{\rm Tr}(e^{-\beta\mathcal{H}(t_{\rm f})}\Phi_{t_{\rm f}}[e^{\beta\mathcal{H}(t_{\rm i})}\rho_{\rm i}])=\ln\left\{\frac{Z_{\rm f}}{Z_{\rm i}}\,{\rm Tr}(\rho_{\rm f}^{\rm th}\Phi_{t_{\rm f}}[(\rho_{\rm i}^{\rm th})^{-1}\rho_{\rm i}])\right\}\nonumber \\
&=& -\beta\Delta F + \ln\{{\rm Tr}((\rho_{\rm i}^{\rm th})^{-1}\rho_{\rm i})\,{\rm Tr}(\rho_{\rm f}^{\rm th}\Phi_{t_{\rm f}}[\widetilde{\rho_{\rm i}}])\} =
-\beta\Delta F + \ln{\rm Tr}((\rho_{\rm i}^{\rm th})^{-1}\rho_{\rm i}) + \ln{\rm Tr}(\rho_{\rm f}^{\rm th}\Phi_{t_{\rm f}}[\widetilde{\rho_{\rm i}}]) \ ,
\end{eqnarray}
where $\widetilde{\rho_{\rm i}}$ is defined as
\begin{equation}
    \widetilde{\rho}_{\rm i} \equiv \frac{(\rho_{\rm i}^{\rm th})^{-1}\rho_{\rm i}}{{\rm Tr}((\rho_{\rm i}^{\rm th})^{-1}\rho_{\rm i})}.
\end{equation}
This immediately leads to
\begin{equation}\label{eq:app_phys_int_1}
    \frac{\mathcal{G}(i\beta)}{\mathcal{G}_{\rm TPM}(i\beta)} = 
    \frac{{\rm Tr}\left(\rho_{\rm f}^{\rm th}\Phi_{t_{\rm f}}[\rho_{\rm i}]\right)}{{\rm Tr}\left(\rho_{\rm f}^{\rm th}\Phi_{t_{\rm f}}[\widetilde{\rho}_{\rm i}]\right)} \ , 
\end{equation}
so that, when $\rho_{\rm i}=\rho_{\rm i}^{\rm th}+\chi$, one gets
\begin{equation}\label{eq:app_phys_int_2}
    \frac{\mathcal{G}(i\beta)}{\mathcal{G}_{\rm TPM}(i\beta)} = d\,\frac{{\rm Tr}\left(\rho_{\rm f}^{\rm th}\Phi_{t_{\rm f}}[\rho_{\rm i}^{\rm th}]\right)+{\rm Tr}\left(\rho_{\rm f}^{\rm th}\Phi_{t_{\rm f}}[\chi]\right)}{{\rm Tr}\left(\rho_{\rm f}^{\rm th}\Phi_{t_{\rm f}}[\mathbb{I}]\right)} \ ,
\end{equation}
with $d$ dimension of the Hilbert space associated to the quantum system.
The reader can find the discussion about the physical interpretation
of Eqs.\,(\ref{eq:app_phys_int_1}) and (\ref{eq:app_phys_int_2}) in the main text.  

\section{Three-level driven system in contact with thermal reservoirs and comparison
  between EPM and MLL schemes}
In this section we summarize the details of the three-level quantum system in contact
with three thermal reservoirs and externally driven, addresed in the main text. The Hamiltonian of the three-level system is written as
\begin{equation}
    H=\omega_3|\epsilon_B\rangle\!\langle\epsilon_B|+\omega_1 |\epsilon_A\rangle\!\langle\epsilon_A| \ ,
\end{equation}
with its eigensystem $\{\ket{\epsilon_g},\ket{\epsilon_A},\ket{\epsilon_B};0,\omega_1,\omega_3\}$.
The external driving term is represented by the following time-dependent Hamiltonian term
\begin{equation}\label{SMdrive}
    H_{\rm{drive}}(t)=g(t)(|\epsilon_g\rangle\!\langle\epsilon_B|+\rm{h.c.})+f(t)(|\epsilon_A\rangle\!\langle\epsilon_B|+\rm{h.c.}).
\end{equation}
driving transitions between the the second excited state and both the ground and first-excited
states. The interaction with the three thermal reservoirs renders
the dynamics of the system open and described to a good approximation via the
Markovian master equation
\begin{equation}\label{me3}
    \dot{\rho}=-i[H+H_{\rm{drive}}(t),\rho(t)]+\sum_{i,j}L_{ij}\rho L_{ij}^\dag-\frac{1}{2}\{L_{ij}^\dag L_{ij},\rho\} \ ,
\end{equation}
where $L_{ij}\equiv\sqrt{\eta_{ij}}|\epsilon_i\rangle\!\langle\epsilon_j|$ and
$\eta_{ji}$ are transition rate. In particular
\begin{equation}
\begin{aligned}
    & \eta_{gA}=\gamma (n_1^{th}+1),\qquad \eta_{Ag}=\gamma n_1^{th},\\
    & \eta_{AB}=\gamma (n_2^{th}+1),\qquad \eta_{BA}=\gamma n_2^{th},\\ 
    & \eta_{gB}=\gamma (n_3^{th}+1),\qquad \eta_{Bg}=\gamma n_3^{th} \ ,
\end{aligned}
\end{equation}
where $n_r^{\rm th}=(e^{\beta_r\omega_r}+1)^{-1}$ and $\omega_3=\omega_2+\omega_1$.

It is easy to see that, choosing $\beta_r=\beta$ $\forall r$
(i.e., there is only one temperature for the environment) and in the absence of external
driving, the thermal state $\rho_{\rm th}=e^{-\beta H}/Z$ is a fixed point of the open dynamics.
The dynamics of the open quantum system without the external driving and with possibly
different temperatures, describes processes involving only heat exchanges which are the main
focus of Ref.~\cite{MicadeiPRL2020}. We consider here this case in order to highlight some of the
differences between our protocol and the TPM and MLL schemes. We refer to the main text for
the results obtained by the numerical analysis of the case in which also the driving term
Eq.~\eqref{SMdrive} is included.
\begin{figure}[t]
\centering
\includegraphics[width=0.6\textwidth]{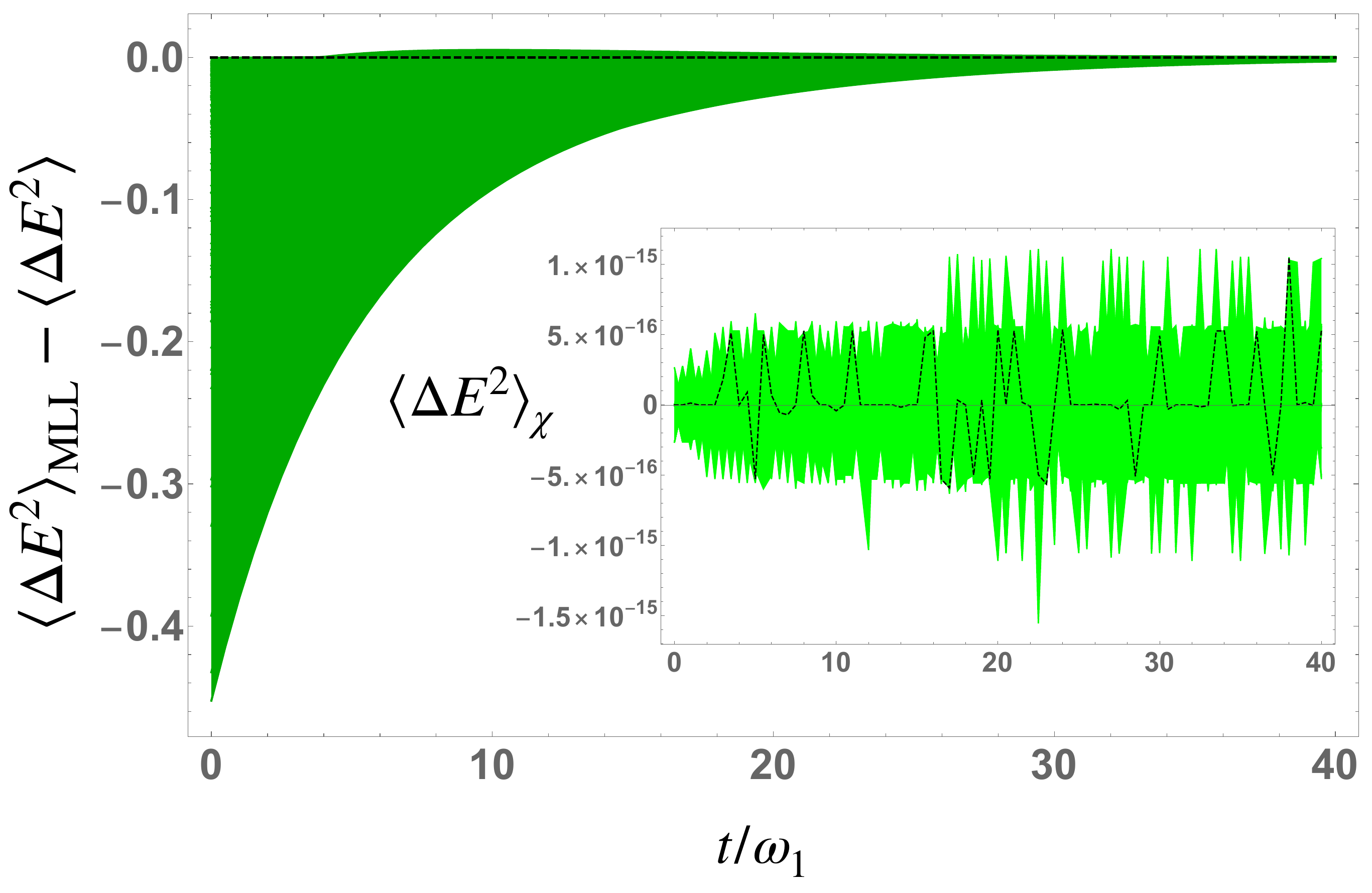}
\caption{Difference between the second moment of
  the MLL probability distribution and the one from the EPM protocol for 100
  randomly sampled initial states. The shaded area comprises all the differences,
  as a function of time, for each initial state. It can be easily seen that only
  seldom the second moment of the MLL scheme results bigger than the one of our scheme.
  It should be noted that, asymptotically the difference vanishes. The inset shows
  that the coherence contribution to the second moment of our distribution
  ($\langle\Delta E^2\rangle_\chi$, obtained from the taking the difference between
  the left hand side of Eq.\eqref{division} and the first term on its right hand side) is,
  in this case, negligible throughout the dynamics. The black dashed curve
  is one instance for a randomly picked initial state (colors online).}
  \label{figsec}
\end{figure}

In Fig.~\ref{figsec}, we show the difference between the second moments of the MLL
and the present EPM protocol probability distributions, as a function of time and for
100 randomly chosen initial state. The inset shows that the coherence contribution is
negligible. It is easy to see that, for the vast majority of cases, the second moment
from our protocol is greater than the one of the MLL scheme. When this happens, we can
already conclude that, in our protocol, we need to pay the freedom deriving from not
performing any initial measurements with an increase in the uncertainty of the probability
distribution. In the few cases, and instants of time, in which the hierarchy of the
second moments is reversed, we need to resort to a more refined notion of uncertainty.
We do so in Fig.~\ref{figentr}, where it is shown
the difference between the Shannon entropy $H$ of our protocol probability distribution
with the one of the MLL and TPM schemes, for the same random sampling of 100 initial states
as before. We see that the Shannon entropy of our protocol is always greater than the one
of the other schemes, which proves the increase of uncertainty due to the initial virtual
measurement. While this result is expected for the comparison between the EPM and MLL schemes,
as discussed before, the comparison with the TPM scheme is consistent with the fact that
the effect of the initial coherence is negligible in this case.
We refer to the numerical section in the main text for a more general case
(see Fig.~\ref{fig:2nd_moment_coh} (b) ). Finally, it should be noted that both the
differences of second moments and Shannon entropies vanish at long times, consistently
with the fact that the system reaches asymptotically a (non-equilibrium) steady-state,
where all the probability distributions introduced before coincide.   

\begin{figure}[ht]
\centering
\includegraphics[width=0.52\textwidth]{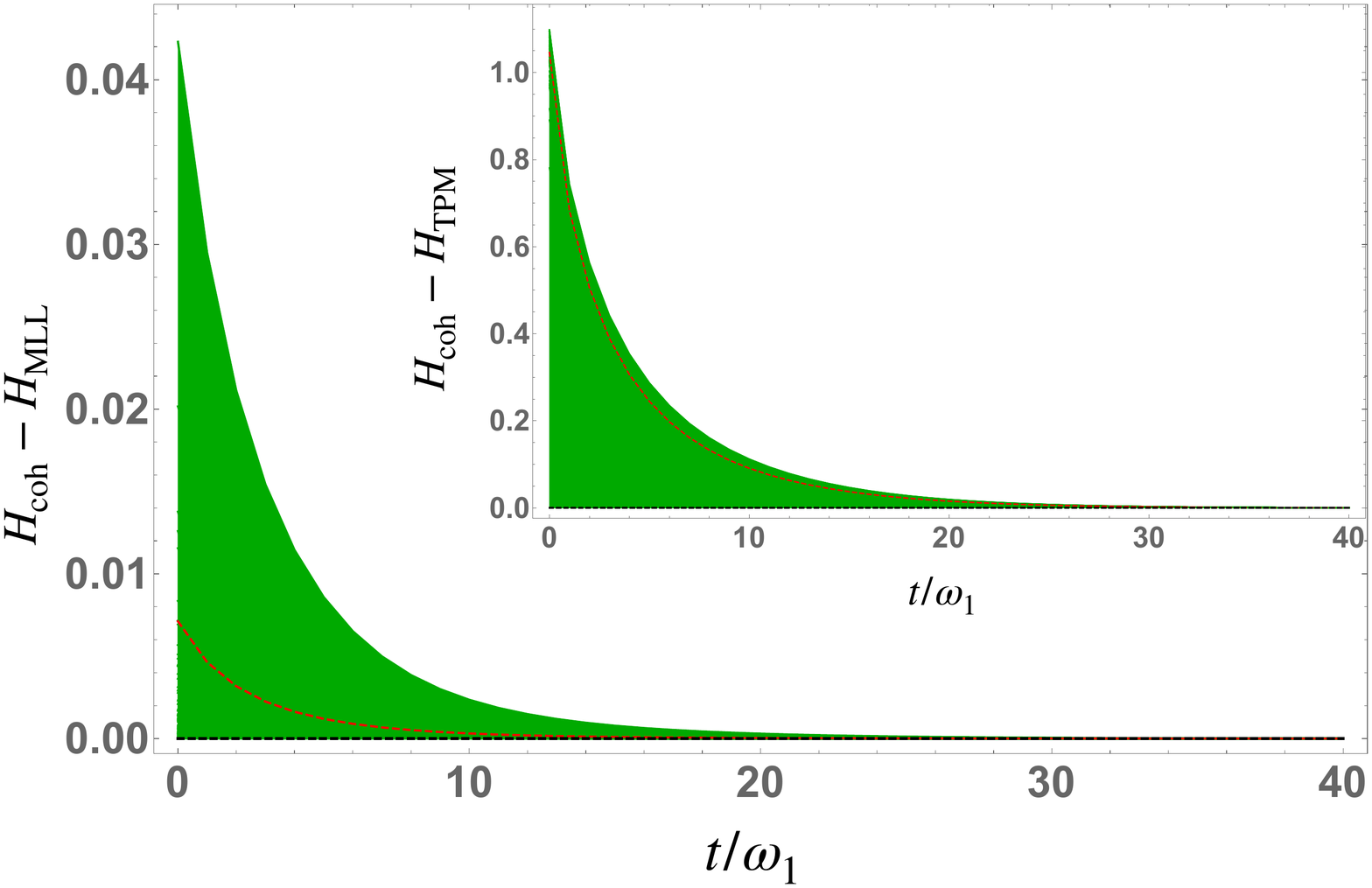}
\caption{Difference between the Shannon entropy of our protocol probability distribution
  and the one of the MLL scheme for 100 randomly sampled initial states. The shaded area
  comprises all the differences, as a function of time, for each initial state. Instead,
  the inset shows the difference between the Shannon entropy of our protocol
  probability distribution and the one of the TPM scheme for the same 100 randomly
  sampled initial states. The red dashed curve is an instance of the differences
  for a randomly picked initial state. It should be noted that, asymptotically,
  the difference vanishes (colors online).}
  \label{figentr}
\end{figure}

\end{widetext}

\bibliography{biblio_AT}

\end{document}